\documentclass[aps,prl,showpacs,twocolumn,superscriptaddress]{revtex4}
\usepackage{graphicx}
\usepackage{amsmath}%
\def\be{\begin{equation}}
\def\ee{\end{equation}}
\def\bea{\begin{eqnarray}}
\def\eea{\end{eqnarray}}

\def\EF{$E_{\rm F} $}

\def\a{$\alpha $}
\def\deg{$^{\circ}$}
\def\A-1{$\AA^{-1}$}
\def\LPB{Li$_{0.9}$Mo$_6$O$_{17} $}
\def\~{$\approx$}

\begin{document}

\title{New Luttinger liquid physics from photoemission on Li$_{0.9}$Mo$_6$O$_{17}$}

\author{Feng Wang}
\affiliation{Randall Laboratory of Physics, University of
Michigan, Ann Arbor, Michigan 48109, USA}
\author{J. V. Alvarez}
\altaffiliation{This work started while J. V. A. was at UM}
\affiliation{Departamento de F\'{\i}sica de la Materia Condensada,
Universidad Autonoma de Madrid, 28049 Madrid, Spain}
\author{S.-K. Mo}
\affiliation{Randall Laboratory of Physics, University of
Michigan, Ann Arbor, Michigan 48109, USA}
\author{J. W. Allen}
\affiliation{Randall Laboratory of Physics, University of
Michigan, Ann Arbor, Michigan 48109, USA}
\author{G.-H. Gweon}
\affiliation{Department of Physics, University of California,
Berkeley, California 94720, USA}
\author{J. He}
\affiliation{Department of Physics and Astronomy, University of
Tennessee, Knoxville, Tennessee 37996, USA}
\author{R. Jin}
\affiliation{Solid State Division, Oak Ridge National Laboratory,
Oak Ridge, Tennessee 37831, USA}
\author{D. Mandrus}
\affiliation{Department of Physics and Astronomy, University of
Tennessee, Knoxville, Tennessee 37996, USA} \affiliation{Solid
State Division, Oak Ridge National Laboratory, Oak Ridge,
Tennessee 37831, USA}
\author{H. H\"ochst}
\affiliation{Synchrotron Radiation Center, University of
Wisconsin, Stoughton, Wisconsin 53589, USA}

\date{\today}

\begin{abstract}
Temperature dependent high resolution photoemission spectra of
quasi-1 dimensional Li$_{0.9}$Mo$_6$O$_{17}$ evince a strong
renormalization of its Luttinger liquid density-of-states
anomalous exponent. We trace this new effect to interacting charge
neutral critical modes that emerge naturally from the two-band
nature of the material. Li$_{0.9}$Mo$_6$O$_{17}$ is shown thereby
to be a paradigm material that is capable of revealing new
Luttinger physics.

\end{abstract}
\pacs{71.10.Pm, 71.10.Hf, 79.60.-i}
\maketitle

Condensed matter physics owes much to research on paradigm
materials, i.e. materials that typify a basic phenomenon and often
a whole class of materials.  The best paradigm materials not only
provide a proof of existence in nature, but also reveal new
physics of the phenomenon.  E.g., research on silicon has shaped
the understanding of covalently bonded semiconductors. This paper
reports research on \LPB\ that not only solidifies its status as a
paradigm quasi-1 dimensional (Q1D) material showing Luttinger
liquid (LL) physics, but goes beyond previous work on this or any
other Q1D material to reveal new LL physics as well.

The LL theory~\cite{Haldane81} provides an understanding of the 1D
orthogonality catastrophe, in which single-electron behavior
cannot be adiabatically continued from a 1D system of free
electrons when electron-electron interactions are included.  This
striking phenomenon is directly observable in angle integrated
photoemission experiments~\cite{Dardel91} as a nearly complete
power law suppression of the single particle density of states
(DOS) at the Fermi energy \EF. LL theory also characterizes (1)
the low-energy excitation spectrum as consisting entirely of
independent, linearly dispersing spin and charge density
fluctuations (spinons and holons, respectively) and (2) the
various correlation functions as displaying power law behaviors
with anomalous exponents, e.g. \a\ for the single particle DOS
near \EF. For this theory to apply in a real solid material, it is
essential to have (i) high anisotropy and (ii) linearity of the
dispersion of the underlying bands over a substantial energy range
away from \EF. The importance of (ii), e.g.~in setting the energy
scale for observing LL behavior, tends to be overlooked.

Among all the few Q1D solids~\cite{otherLL} for which LL behaviors
are claimed, \LPB\ is unique as a metal with a well understood
band structure~\cite{Whangbo88} that clearly satisfies both
criteria. Strong anisotropy has been confirmed by various
experiments~\cite{Greenblatt84,Choi04} and linear dispersion over
a range $\approx0.13$ eV from \EF\ is observed in angle resolved
photoemission spectroscopy
(ARPES)~\cite{JDD99,Gweon01,Gweon02,Gweon03,Allen02,Gweon04},
which directly measures the single particle spectral density as a
function of momentum ${\it k}$ and energy $\omega$.

ARPES spectra~\cite{JDD99,Gweon01,Gweon02,Gweon03,Gweon04,Allen02}
have shown expected holon and spinon~\cite{Gweon03} features
near \EF, and also a power law DOS when angle integrated.
These findings are strongly supported
qualitatively by recent tunneling spectroscopy~\cite{Hager05} that
also shows a power law DOS near \EF.  However, the values of \a,
$\approx 0.6$ and $\approx0.9$, extracted from tunneling and
ARPES, respectively, are quite different.  The only hope of
reconciling this difference lies in the differing T ranges, 5-55
K, and 250-300 K of the two experiments, respectively.
Unfortunately, as we show below, such a large T-renormalization of
\a\ is inconsistent with \LPB\ transport properties implying
T-independent \a\, within usual one-band LL theory.  In this
Letter we report new ARPES directly showing that the large
T-renormalization of \a\ does indeed occur and we show that its
origin lies in interacting charge neutral critical modes that
emerge naturally for \LPB\ because two bands cross \EF.  The
theory also predicts the possibility of observing two distinct
kinds of \a(T) behavior on various sample surfaces, as may
actually have been observed.  This is both the first observation
and the first theory of such a strong anomalous exponent
T-dependence.

\begin{figure}[t]
\vspace{-0.03in}
\includegraphics[width=3.1 in]{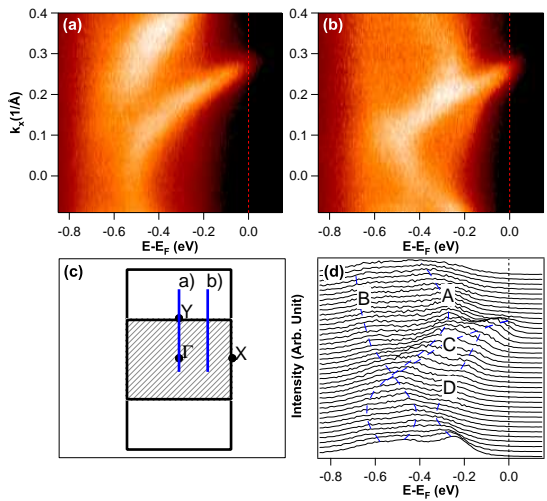}
\caption{ARPES spectra of \LPB\ measured at T = 200 K, $h\nu$ = 30
eV. (a), (b): Energy versus momentum spectra taken along
$\Gamma-Y$ and off-normal direction, respectively. (c): A
schematic view of the 1-D FS with the k space paths of (a) and (b)
indicated. (d): Stack view of (b), with dashed lines highlighting
the band dispersions.} \vspace{-0.15in} \label{BandStructure}
\end{figure}

ARPES experiments over a wide T-range 15-300 K, with photon
energies 24 eV and 30 eV, were performed at the U1-NIM undulator
beamline of the Synchrotron Radiation Center of the University of
Wisconsin. Single crystal samples were grown by the temperature
gradient flux method and were cleaved {\it in-situ} at pressure $<
1 \times 10^{-10}$ torr for measurement. A Scienta SES2002
spectrometer with an acceptance angle of 14\deg\ was used for all
data. Freshly evaporated gold was measured before and after the
experiment at temperature  $\lesssim15$ K, for a careful
calibration of the Fermi enegy as well as the instrumental energy
resolution, $\Delta E\approx18$ meV FWHM\@. We have collected
extensive data for samples cleaved at different temperatures and
with the measurement temperature both increasing and decreasing.

The ARPES spectra in Figs. 1(a) and 1(b) were taken with angular
resolution $\approx0.3$\deg\ and the analyzer slit perpendicular
to the Q1D Fermi surface (FS) to show the band dispersions. The
associated k-space paths are illustrated in Fig. 1(c). \LPB\ has 4
Mo 4{\it d} bands near \EF, labeled A - D in Fig. 1(d). Two of
them (C and D) become degenerate at binding energy $\approx0.13$
eV, then disperse linearly~\cite{Gweon03} and cross \EF\ together
to define the Q1D FS. Along path (a), the intensity of band D is
very suppressed by the ARPES matrix element. However, both C and D
are clearly seen along path (b). The measured band structure is in
good agreement with that found previously by ARPES and
qualitatively with the band structure calculations. The same
spectra were taken at various temperatures down to 15 K and show
no apparent change of the bands.

\begin{figure}[t]
\vspace{-0.06in}
\includegraphics[width=3.3 in]{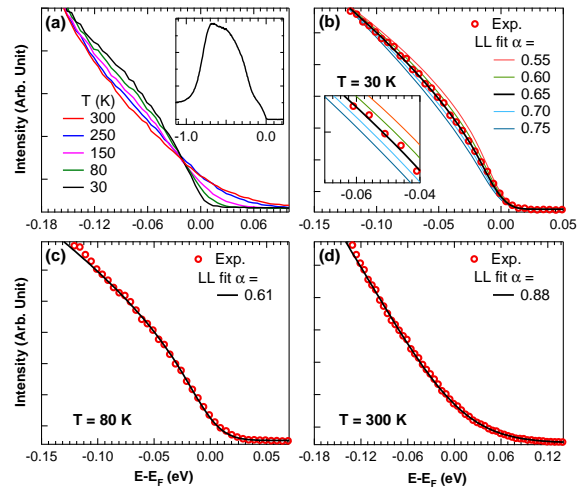}
\caption{(a): T-dependent spectra angle integrated along path for
Fig. 1(a). Only selected temperatures are shown for clarity. Inset
shows the wide energy range spectrum at 30K. (b), (c), (d):
Fitting with finite T LL spectral function. Inset in (b) shows the
details of the fitting between 0.04-0.07 eV.} \vspace{-0.15in}
\label{Fitting}
\end{figure}

Angle integrated spectra were taken for the same geometries as for
Figs. 1(a) and 1(b). Fig. 2(a) shows a set of spectra for selected
temperatures, with the inset showing a wide range spectrum for 30
K. All spectra are normalized to the intensity between binding
energy 0.2-0.8 eV, which is mainly from bands A and B in Fig. 1.
At all T the power law suppression of the near \EF\ DOS is
evident. As T decreases there is much sharpening, a great
reduction of the tail above \EF\ and a strong increase of the
curvature. This sharpening is only partly a reduction of thermal
broadening. There is also a significant decrease of the power law
exponent \a\, as we have confirmed and quantified by comparing the
experimental spectra with the LL DOS using the finite temperature
theory of Ref.~\cite{Orgad01}. The calculated theory spectra were
convolved with a Gaussian with the width set by the instrumental
energy resolution $\Delta E$, and were scaled as needed because
the relative experimental intensities for various T are not well
calibrated.  As illustrated in Figs. 2(b)-(d) the experimental
spectral shape at each T can be very well fit by the theory up to
binding energy 0.105 eV, above which the A band weight starts to
come in. The \a\ value is T-dependent as plotted in Fig. 3(a).
Consistent with previous photoemission and tunneling, \a\ starts
from $\approx0.9$ at room temperature, decreases sharply between
$300$ K and $150$ K, then slows down and saturates at $\approx0.6$
between $150-50$ K.  At even lower temperature, there is a slight
revival of \a, as documented in Fig. 2,(b) and (c). Over the whole
range, \a\ is never less than 0.6. Fig. 3(a) also shows \a(T)
derived from spectra taken with the geometry of Fig. 1(b).

Our interpretation of this \a(T) behavior, which yields the solid
theory curve of Fig. 3(a),  flows from the excellent agreement,
found here and previously~\cite{Gweon03,Gweon04}, between finite-T
LL~\cite{Orgad01} spectral theory and measured photoemission
spectra in the full range of T over which \a\ (which itself is
very large) renormalizes. Therefore, the most suitable starting
point for our calculation is an effective theory based on the LL
where T establishes the low-energy cutoff. However, in a
single-band spin rotational invariant LL any variation of \a\
comes from the renormalization of $K_{\rho}$, the Luttinger
parameter in the charged sector through the relation $\alpha
=(K_{\rho}+K_{\rho}^{-1}-2)/4$.  The driving force of that
renormalization may be (i) Umklapp scattering, present only when
$4k_F=G$ where $k_F$ is the Fermi vector and $G$ is a reciprocal
lattice vector or (ii) for an incommensurate system, the interplay
of weak disorder and backward scattering, which recouples charge
and spin~\cite{Giamarchi88}. Case (i) is forbidden here because
$k_F$ in \LPB\ is incommensurate~\cite{Whangbo88}. Further, in
both cases any temperature change in \a\ is necessarily caused by
renormalization in the charged sector and so must be correlated
with strong transport anomalies in exactly the same range of
temperatures. In contrast, we have found that the resistivity for
\LPB\ displays a nearly perfect power-law in T for almost one
decade~\cite{footnote}. This lack of scaling relation between the
spectral and transport exponents suggests that the renormalization
process is taking place in a neutral sector. Both our
photoemission experiments and band theory
calculations~\cite{Whangbo88} show two bands crossing the Fermi
level, which offers a natural origin for such neutral modes as the
fluctuations in the difference of density between the two bands
$\rho_{-}(x)=(\rho(x)_{C \uparrow}+\rho(x)_{C \downarrow})
-(\rho(x)_{D \uparrow}+\rho(x)_{D \downarrow})$. Following the
standard bosonization process we can decouple the fixed-point
Hamiltonian in the form:
\begin{equation}
H= \sum_{\substack{\alpha=\rho,\sigma \\ \beta=\pm}}
\frac{v_{\alpha \beta}}{2} \int dx \left[K_{\alpha \beta}
(\Pi(x)_{\alpha \beta})^2+ \frac{1}{K_{\alpha \beta}} (\partial_x
\phi(x)_{\alpha \beta})^2 \right] \nonumber
\end{equation}
where  $\Pi(x)_{\alpha \beta}$ and $\phi(x)_{\alpha \beta}$ are
the conjugate bosonic fields, $v_{\alpha \beta}$ the sound
velocities and $K_{\alpha \beta}$ the Luttinger
parameters($\beta=+$/$-$ superposition/difference,
$\alpha=\rho$/$\sigma$ charge/spin density). The $\rho_{-}$ and
$\sigma_{-}$ modes can be pictured as the 1D-allowed bosonic
excitations into which an interband exciton can decay. Having two
extra modes with different velocities, we expect wide ARPES
lineshapes as has actually been observed in \LPB~\cite{Gweon03},
and now $\alpha
=(K_{\rho+}+K_{\rho+}^{-1}+K_{\rho-}+K_{\rho-}^{-1}-4)/8$.

\begin{figure}[t]
\vspace{-0.04in}
\includegraphics[width=3.3 in]{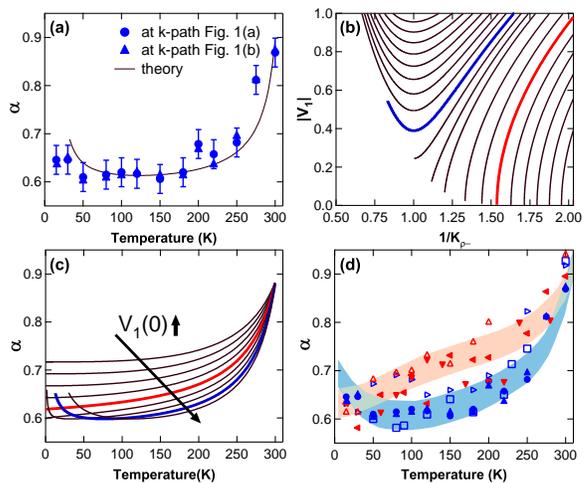}

\caption{(a): \a(T) extracted from fitting of the spectra in Fig.
2(a)(circles) for a complete T series and the same (triangles)
measured for the k-path of Fig. 1(b) on the same sample. Typical
error bar from fitting $\approx\pm0.03$. The line is a theory
calculation with our model (see text). (b): Kosterlitz-Thouless
flow for $V_{2}=0,g_{\sigma-}=0$;  finite values give the same
basic structure. (c): \a(T) as $V_{1}$
increases, showing two different behaviors. (d): \a(T) measured on
6 samples, with error bars (not shown) similar to (a). Shading
highlights two different behaviors much like those in theory curves of
panel (c)(see text).}

\vspace{-0.15in} \label{alpha_T}
\end{figure}

Let us consider now how these critical modes can interact via
marginal interactions (those that can renormalize critical
exponents). The interactions compatible with the symmetries
represent intra and inter-band spin-backscattering and
pair-tunneling processes between the two bands.  A convenient form
in which to write the renormalization group equations is:
\begin{eqnarray}
(K_{\rho -})^{\prime } &=&\frac{1}{4}\left\{3(V_{2})^2+V_{1}^2\right\} \;,  \label{E1} \\
(V_{1})^{\prime } &=& \left\{ d+g_{\sigma +}\right\}V_{1}+g_{\sigma -}V_{2} \;,  \label{E2} \\
(V_{2})^{\prime } &=& g_{\sigma -}V_{1}+\left\{d-g_{\sigma+}\right\}V_{2} \;,\label{E3} \\
(g_{\sigma+})^{\prime } &=&-\left\{ g_{\sigma -}^2+g_{\sigma
+}^2\right\}-
 \frac{1}{2}(V_{2})^{2} \;, \label{E4} \\
(g_{\sigma-})^{\prime } &=&-2\left\{g_{\sigma -} g_{\sigma
+}\right\}+\frac{1}{2}V_{2} V_{1}\; \label{E5}
\end{eqnarray}
$g_{\sigma\pm}$ represent spin-backscattering process, $V_{i}$
interband pair-tunneling and $d=1-\frac{1}{K_{\rho -}}-g_{\sigma
+}$.

Since we have restricted ourselves to spin rotational invariant
flows, $K_{\sigma +}$ and $K_{\sigma -}$ do not show up explicitly
in the equations. The total charge mode ($\rho +$) contributes to
the total value of $\alpha$ but the absence of Umklapp scattering
keeps the charged sector independent and free of significant
renormalizations, consistent with the T-independence of the
resistivity exponent in this range of T, and with the low T
saturation value of \a(T) being greater than 0.6.  The lack of
commensurability has been confirmed by photoemission and the two
band structure calculations available and reinforces the essential
role of the neutral mode for \a(T).  The interchain 3D
perturbations do not seem to play any important role at these
temperatures, the only evidence of the 3D physics being the
superconducting transition at 1.8
K~\cite{Greenblatt84,Schlenker85,footnote}.

We are not interested here in the correlations dominating at T = 0
which have already been studied in detail (see for example
Ref.~\cite{Wu03} and references therein). Rather we focus on the
crossover structure at higher temperatures which in \LPB\ has
great physical significance due to the large energy range over
which the bands crossing \EF\ have linear dispersion, as pointed
out above.
Fig. \ref{alpha_T}(b) shows the result of integrating the RG
equations for varying $V_{1}(0)$. For $V_{2}(0)=g_{\sigma-}(0)=0$
and $g_{\sigma+} > 0$, Eqs. (\ref{E1}) and (\ref{E2}) display
literally a Kosterlitz-Thouless \cite{KT73} (KT) flow in the
charge density difference ($\rho-$), with decoupled gapless
excitations for the total spin ($\sigma +$). This behavior is not
isolated.  To be stable in the range of temperature of interest,
finite values of $V_{2}(0)$ and positive values $0<g_{\sigma-}(0)
< g_{\sigma+}(0)$, it is sufficient that
$\lambda=d+\sqrt{g_{\sigma +}^2+g_{\sigma -}^2}<0$. Besides
capturing the LL behavior and the instabilities of the two-band
system, this KT flow supports simultaneously the correct  magnetic
(Pauli susceptibility \cite{Greenblatt84,Choi04}) and spectral (T
dependence of \a) phenomenologies. Fig. \ref{alpha_T}(c) shows the
\a(T) corresponding to the flows of Fig. \ref{alpha_T}(b).  The
theory curve of Fig. \ref{alpha_T}(a) gives a good general
description of the data, thus placing \LPB\ in a region of
parameters where the RG flow and the \a(T) can be understood in
terms of the KT behavior. i) The downward renormalization of
$\alpha$ comes from a reduction of the $1/K_{\rho-}$ and a
simultaneous screening of the V couplings. ii) At small values of
$V_{1,2}$ and specially, close to the separatrix ending in
$K_{\rho-} \sim 1$, the change in $1/K_{\rho-}$ is very slow (see
Eq. (\ref{E1})). As a consequence, $\alpha$ displays a saturation
regime over a large range of intermediate temperatures. iii) The
incipient revival of \a\ takes place when $K_{\rho-}=1$ and at
lower temperatures the V's flow to strong coupling. The revival
takes place at the lowest temperatures we could measure so we
cannot determine experimentally if it culminates in a gap opening
in the $\rho-$ sector. If so, that crossover and the eventual gap
opening might help to understand  the broad anomaly in the
specific heat~\cite{Schlenker85,Choi04} that is present in the
same range of T.

The proximity of \LPB\ to a separatrix suggests the possibility
that both behaviors of \a(T) seen in Fig. \ref{alpha_T}(c) might
be observed in experiment.  As $1/K_{\rho-}$ decreases the {\em
electronic} compressibility increases.  Therefore the $\rho_{-}$
mode tends to favor phase separation, perhaps nucleated by defects
in different parts of the sample. In the first ``phase'' the
reduction of stiffness is large enough to support a revival of \a.
On the other side of the separatrix, the change of \a\ is slower
and monotonic.  In fact, both behaviors may have been observed.
Fig. \ref{alpha_T}(d) plots the \a(T) measured for 6 different
sample surfaces. Each symbol corresponds to one sample except for
the data repeated from Fig. 3(a). The solid symbols represent
situations where the sample was cleaved at high T and T was then
decreased, and the open symbols represents situations where the
sample was cleaved at 50 K and T was then increased after quickly
visiting the lowest measured T. Despite slight differences from
sample to sample, our main finding of a decrease and flattening of
\a\ with decreasing T is observed robustly in all 6 samples.
However, as illustrated by the shadings in the figure, it appears
that the samples can be separated into two groups according to the
rapidity of the decrease of \a(T). The \a(T) of one group
decreases faster and flattens at a higher temperature than occurs
for the other group. Moreover, the first group tends to show the
slight revival of \a(T) at low temperature while the slower group
does not, generally consistent with Fig. \ref{alpha_T}(c). Any
variations seen in individual data sets beyond these general
features are deemed to lie within the typical error bars (not
shown but like those in Fig. \ref{alpha_T}(a)). Further work will
be needed to fully characterize the two behaviors.

In summary, T-dependent photoemission of \LPB\ shows that the
anomalous exponent \a\ has a large renormalization with decreasing
T, nicely connecting the high and low T values found previously in
photoemission and tunneling, respectively.  New theory reveals the
essential role of neutral charge modes present specifically
because there are two bands crossing \EF, the absence of which
would make the observed \a(T) impossible to rationalize with a
resistivity T-dependence that shows no exponent renormalization.
The possibility of two distinct \a(T) behaviors, as may have been
seen experimentally, is also intrinsic in the theory.  This is
both the first observation of such a strong T-renormalization of
\a(T) and the first theory of this unexpected behavior. Moreover,
the case that \LPB\ is the long sought paradigm of LL behavior in
a solid material has thereby been brought to a new level.
\begin{acknowledgments}
This work was supported at UM by the U.S. NSF (DMR-03-02825), at
UAM by MEC under a contract RyC, at the ORNL by the U.S. DoE
(DE-AC05-00OR22725), at UT by the U.S. NSF (DMR-00-72998), at the
SRC by the U.S. NSF (DMR-00-84402).

\end{acknowledgments}



\begin{thebibliography}{10}

\bibitem{Haldane81} F. D. M. Haldane, J. Phys. C {\bf 14}, 2585 (1981).

\bibitem{Dardel91} B. Dardel {\it et al.}, Phys. Rev. Lett. {\bf
67}, 3144 (1991).

\bibitem{otherLL} E.g., A. Schwartz {\it et al.}, Phys. Rev. B {\bf 58}, 1261
(1998); R. Claessen {\it et al.}, Phys. Rev. Lett. {\bf 88},
096402 (2002).

\bibitem{Whangbo88} M.-H. Whangbo and E. Canadell, J. Am. Chem. Soc. {\bf 110}, 358 (1988).

\bibitem{Greenblatt84} M. Greenblatt  {\it et al.}, Solid State Comm. {\bf 51}, 671 (1984).

\bibitem{Choi04} J. Choi {\it et al.}, Phys. Rev. B {\bf 69}, 085120 (2004).

\bibitem{JDD99} J. D. Denlinger  {\it et al.}, Phys. Rev. Lett. {\bf 82}, 2540 (1999).

\bibitem{Gweon01} G.-H. Gweon {\it et al.}, J. Elec. Spect. Relat. Phenom. {\bf 117-118}, 481 (2001).

\bibitem{Gweon02} G.-H. Gweon {\it et al.}, Physica B {\bf 312-313}, 584 (2002).

\bibitem{Allen02} J. W. Allen, Solid State Comm. {\bf 123}, 469 (2002).

\bibitem{Gweon03} Because \a\ $>$ 0.5 the spinon feature appears as the leading edge in the lineshape.
G.-H. Gweon, J.W. Allen, and J.D. Denlinger, Phys. Rev. B {\bf 68}, 195117 (2003).

\bibitem{Gweon04} G.-H. Gweon {\it et al.}, Phys. Rev. B {\bf 70}, 153103 (2004).

\bibitem{Hager05} J. Hager {\it et al.}, Phys. Rev. Lett. {\bf 95}, 186402 (2005).

\bibitem{Orgad01} D. Orgad, Phil. Mag. B {\bf 81}, 377 (2001).

\bibitem{Giamarchi88} T. Giamarchi and H. J. Schulz,  Phys. Rev. B {\bf 37}, 325 (1988).

\bibitem{footnote} Our analysis of resistivity data, e.g. Ref. [6],
shows positive and negative power laws above and below 25K,
respectively.
This behavior will be discussed from the LL viewpoint in a future
publication. Early thinking that the resistivity increases below
25K signals a density wave transition has been set aside for lack
of any supporting evidence as summarized in Refs. [6, 11, 13].

\bibitem{Schlenker85} C. Schlenker {\it et al.}, Physica {\bf 135B}, 511 (1985).

\bibitem{Wu03} C. Wu, W. V. Liu, and E. Fradkin, Phys. Rev. B { \bf 68}, 115104 (2003).

\bibitem{KT73} J. M. Kosterlitz and D. J. Thouless, J. Phys C {\bf 6}, 1181 (1973).

\end{thebibliography}
\end{document}